\documentclass[aps,prl,twocolumn,english,showpacs,superscriptaddress]{revtex4-1}
\usepackage{graphicx}
\usepackage{amssymb}
\usepackage{amsmath, amstext, amssymb, amsfonts, amsxtra, mathtools}  
\usepackage{braket}
\usepackage{textcomp}
\usepackage{relsize}
\usepackage[usenames,dvipsnames]{color}
\usepackage{units}
\usepackage[normalem]{ulem}

\newcommand{\bonn}{HISKP, University of Bonn, Nussallee 14-16, 53115 Bonn, Germany}
\newcommand{\bonnpi}{Physikalisches Institut, University of Bonn, Nussallee 12, 53115 Bonn, Germany}

\begin{document}
\begin{abstract}
  We propose theoretically how unconventional superconducting pairing in a repulsively interacting Hubbard ladder can be enhanced via the application of a Floquet driving.
Initially the Hubbard ladder is prepared in its charge-density-wave dominated ground state. A periodic Floquet drive is applied which modulates oppositely the energy offset of the two chains and effectively reduces the tunneling along the rungs. This modulation of the energy offsets might be caused by the excitation of a suitable phononic mode in solids or a superlattice modulation in cold atomic gases. We use state-of-the-art density matrix renormalization group methods to monitor the resulting real-time dynamics of the system. We find an enormous enhancement of the unconventional superconducting pair correlations by approximately one order of magnitude. 
\end{abstract}

\title{Dynamically enhanced unconventional superconducting correlations in a Hubbard ladder} 

\author{Ameneh Sheikhan}
\affiliation{\bonn}
\affiliation{\bonnpi}
\author{Corinna Kollath}
\affiliation{\bonn}
\affiliation{\bonnpi}
\maketitle
Recent experimental progress in solid state and quantum gas experiments opened up the possibility to investigate phenomena far-from equilibrium \cite{Orenstein2012, GiannettiMihailovic2016, ZhangAveritt2014, BlochZwerger2008}. 
In pump-probe experiments on solids, the material is brought out-of equilibrium by a short pump laser pulse which either couples mainly to the electronic \cite{Orenstein2012, ZhangAveritt2014, PerfettiWolf2006, SchmittShen2008, HellmannRossnagel2010, RohwerBauer2011, GedikHardy2004, CortesBovensiepen2011, SmallwoodLanzara2012} or structural \cite{RiniCavalleri2007, FoerstCavalleri2011, SubediGeorges2014} modes of the material and the subsequent dynamics is monitored, e.g.~by a probe pulse. The tailored pump pulse can induce a change in the properties of the material and states or properties  which are not accessible in equilibrium can become reachable. Examples of such interesting non-equilibrium states reach from Floquet-Bloch states at the surface of topological insulators \cite{WangGedik2013} over hidden states in  a layered dichalcogenide crystal \cite{StojchevskaMihailovic2014} to states in which the electron-phonon coupling is strongly enhanced \cite{PomaricoGirez2017}. An intensified effort has, in particular, been devoted to dynamically induce superconducting properties in materials \cite{FaustiCavalleri2011, KaiserCavalleri2012, MankowskyCavalleri2014, NicolettiCavalleri2014, CasandrucCavalleri2015, MitranoCavalleri2016}.
However, despite the outstanding experimental progress the understanding of the underlying mechanisms of the dynamic control of materials and their properties such as superconductivity is still far from complete. Simplified models are being considered in order to identify the important ingredients of processes taking place. For example, the enhancement of s-wave superconducting correlations by the excitation of a specific phononic mode has been proposed \cite{SentefKollath2016} or by the use of competing orders\cite{SentefKollath2017}. The build-up of short range Cooper-pair correlations \cite{DasariEckstein2018, SchlawinJaksch2018} is simulated in the driven attractive Hubbard model .

In this work, we show how unconventional superconducting pair correlations can be enhanced drastically by approximately one order of magnitude using tailored light-pulses. We consider a material in a charge density wave ground state which is described by a Hubbard ladder. We assume that a phononic mode can be excited which leads to a modulation of the energy offset between the two legs of the ladder. Alternatively, this situation can be emulated in cold atom experiments using time-dependent superlattices. The periodic modulation effectively lowers the tunneling along the rung of the ladder and  thus drives the system towards the d-wave superconductor.

We consider a material which in equilibrium can be described by a Hubbard ladder with repulsive onsite interactions,  
\begin{eqnarray}\label{eq:Ham_t}
&&H_0(J_\perp) = -J_\parallel \sum_{j,\sigma,m=0,1} \left( c^\dagger_{m,j,\sigma} c_{m,j+1,\sigma}+\mathrm{h.c.}\right)\nonumber\\
&&\qquad\,\,\,-J_\perp \sum_{j,\sigma} \left( c^\dagger_{0,j,\sigma} c_{1,j,\sigma}+\mathrm{h.c.}\right)\nonumber\\
&&\qquad\,\,\,+ U \sum_{j,m=0,1} n_{m,j,\uparrow} n_{m,j,\downarrow}.
\end{eqnarray}
Here, $c_{m,j,\sigma}$ $(c_{m,j,\sigma}^{\dag})$ annihilates (creates) a fermion with spin $\sigma=\uparrow,\downarrow$ on leg $m=0,1$ and rung $j=1,\dots,L$ of the ladder. The operator $n_{m,j,\sigma} = c_{m,j,\sigma}^{\dag} c_{m,j,\sigma} $ is the corresponding density operator. The fermions tunnel along the legs of the ladder with tunneling amplitudes $J_\parallel$  and along the rungs with a tunneling amplitude $J_\perp$. Fermions on the same site interact repulsively with strength $U>0$.  
 The Hubbard ladder is a quasi-one-dimensional model in which some of the two-dimensional characteristics appears. It is one of the rare examples, where a purely repulsive interaction leads to a superconducting state \cite{Giamarchibook}. In the limit of a large tunneling amplitude on the rungs, i.e.~$J_\perp/J_\parallel$ an intuitive picture of the origin of this superconducting phase can be gained. In this limit, the fermions energetically favor to form singlet pairs extended over the rungs. At half filling, each rung is locked into a singlet state. Under doping of the half-filled case by two holes, two situations can occur: First the holes are located on different rungs and two singlets are broken. Second the holes are located on the same rung, only one singlet is destroyed. Since the second configuration of paired holes is energetically favorable, these pairs of holes can Bose-condense under certain conditions. They give rise to a superconducting state with unconventional pairing symmetry. This is a quasi-one dimensional example for the RVB (resonant valence bond) mechanism which was proposed as a possible origin of high-temperature superconductivity \cite{Anderson1997book}.

We assume that by the coupling to a light field, phononic modes can be excited which result in an alternation of the potential underlying the legs of the Hubbard ladder. We describe these phononic modes classically by a time-dependent potential $V(t)$ with 
\begin{eqnarray}\label{eq:V_t}
&&V(t)=\sum_{j,\sigma,m=0,1} A(t) (-1)^m \sin(\omega t) n_{m,j,\sigma}, 
\end{eqnarray}
where $A(t)$ is the amplitude of the modulation which varies slowly in time compared to the oscillation period. Such a potential change could also be emulated in an optical lattice using a superlattice modulation \cite{AidelsburgerBloch2011}. 

We will determine the time-evolution of the model quasi-exactly using the numerical time-dependent DMRG or MPS methods. We show how this modulation can be used starting from a charge density wave to enhance the unconventional superconducting correlations dynamically by an order of magnitude. However, before we do this, we first discuss the ground state properties of the repulsively interacting Hubbard ladder \cite{Giamarchibook}, and then give an insight into the effect of such a modulation deriving a time-independent effective Hamiltonian using the Floquet representation of the system.

The ground state properties of the repulsively interacting Hubbard ladder have been studied both analytically and numerically (see e.g.~\cite{Giamarchibook, DolfiTroyer2015, NoackScalapino1996} and references therein).

At weak interaction, the bosonization method \cite{Giamarchibook} predicts the occurrence of unconventional superconductivity and of a $4k_F$-charge density wave (CDW$^{4k_F}$), where $k_F$ is the Fermi wave-vector. The unconventional superconducting pairs are singlets formed on the rungs of the ladder given by $\Delta_d(j)=c_{0,j,\uparrow}c_{1,j,\downarrow}-c_{0,j,\downarrow}c_{1,j,\uparrow}$ and are often loosely called of d-wave symmetry. At long distances, bosonization predicts that the density and unconventional pair correlations decay algebraically as 
\begin{align}
&\langle \Delta n(j) \Delta n(j+l) \rangle \propto l^ {-K_n}\nonumber\\
&\langle \Delta_d(j) \Delta_d^\dagger(j+l)\rangle \propto l^ {-\frac{1}{K_d}},
\label{eq:correlations}
\end{align}
respectively. Within bosonization the exponents of the decay are related by  $K_\rho=K_n=K_d$ and $K_\rho$ is called the Luttinger liquid parameter. If $K_\rho>1$ the d-wave superconductor is the dominant ground state whereas for $K_\rho<1$ the $4k_F$-charge density wave (CDW$^{4k_F}$) dominates. In contrast, on each leg of the ladder eg. leg with index $m=0$ the spin correlations $\langle S_z(j) S_z(j+l) \rangle$ with $S_z(j)=n_{j,0,\uparrow}-n_{j,0,\downarrow}$ decay exponentially due to the occurrence of a spin gap.

The field-theoretical description has been supported by early numerical work \cite{NoackScalapinto1994,NoackScalapino1995,NoackScalapino1996} which found dominating unconventional pair correlations already on relative short distances and confirmed the existence of an unconventional superconducting state. A decade later in Ref.~\cite{DolfiTroyer2015} the numerical determination of the correlations by DMRG and extrapolation of results have been sufficiently good to confirm the relation between the exponents of the decay of the pair and density correlations (Eq.~\ref{eq:correlations}), i.e.~the relation $K_\rho=K_n=K_d$. Moreover, the Luttinger parameter could be determined reliably for a few specific parameter sets. 

\begin{figure}
\includegraphics[width=0.99\linewidth]{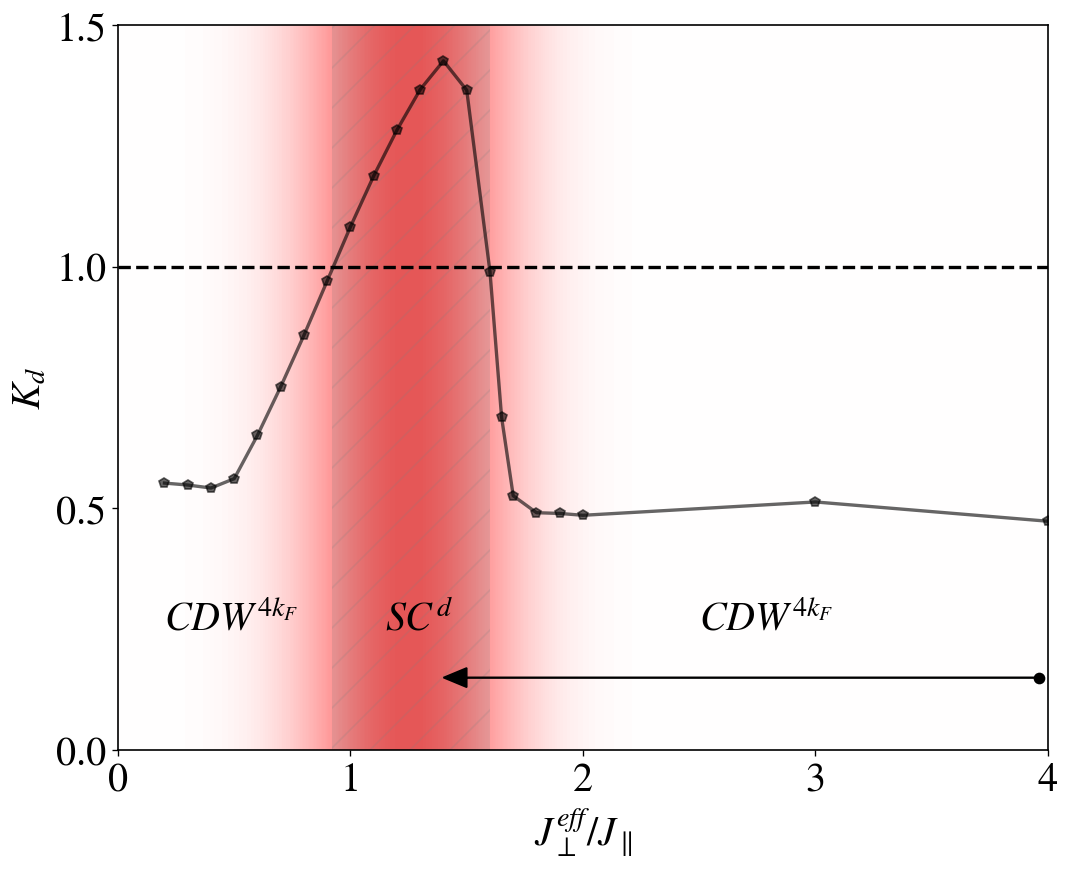}
\caption{\label{fig:Kd_idmrg} Ground state phase diagram of the repulsively interacting Hubbard ladder for $U=8J_\parallel$ at filling $n=0.9375$.  For intermediate hopping values ($0.9\lessapprox J_\perp^\text{eff}/J_\parallel\lessapprox 1.65$) the exponent $K_d>1$ shows that the unconventional pair correlation is dominant, whereas at low and large values of the ratio of the tunneling the charge density wave is dominant since $K_d<1$. The crossover takes place at $K_d\approx 1$ (marked by the dashed horizontal line). $K_d$ is extracted from the decay of the unconventional pair correlations calculated via iDMRG. The unit cell in the iDMRG calculation has size $32\times 2$. The arrow sketches the non-equilibrium situation in the effective Floquet model. Starting with the charge density wave (CDW$^{4k_F}$) with $J_\perp/J_\parallel = 3.9598$, the frequency of the periodic driving is chosen such that the final effective system is an unconventional superconductor (SC$^d$) with $J_\perp^\text{eff}/J_\parallel = 1.4$.} 
\end{figure}

\begin{figure}
\includegraphics[width=0.99\linewidth]{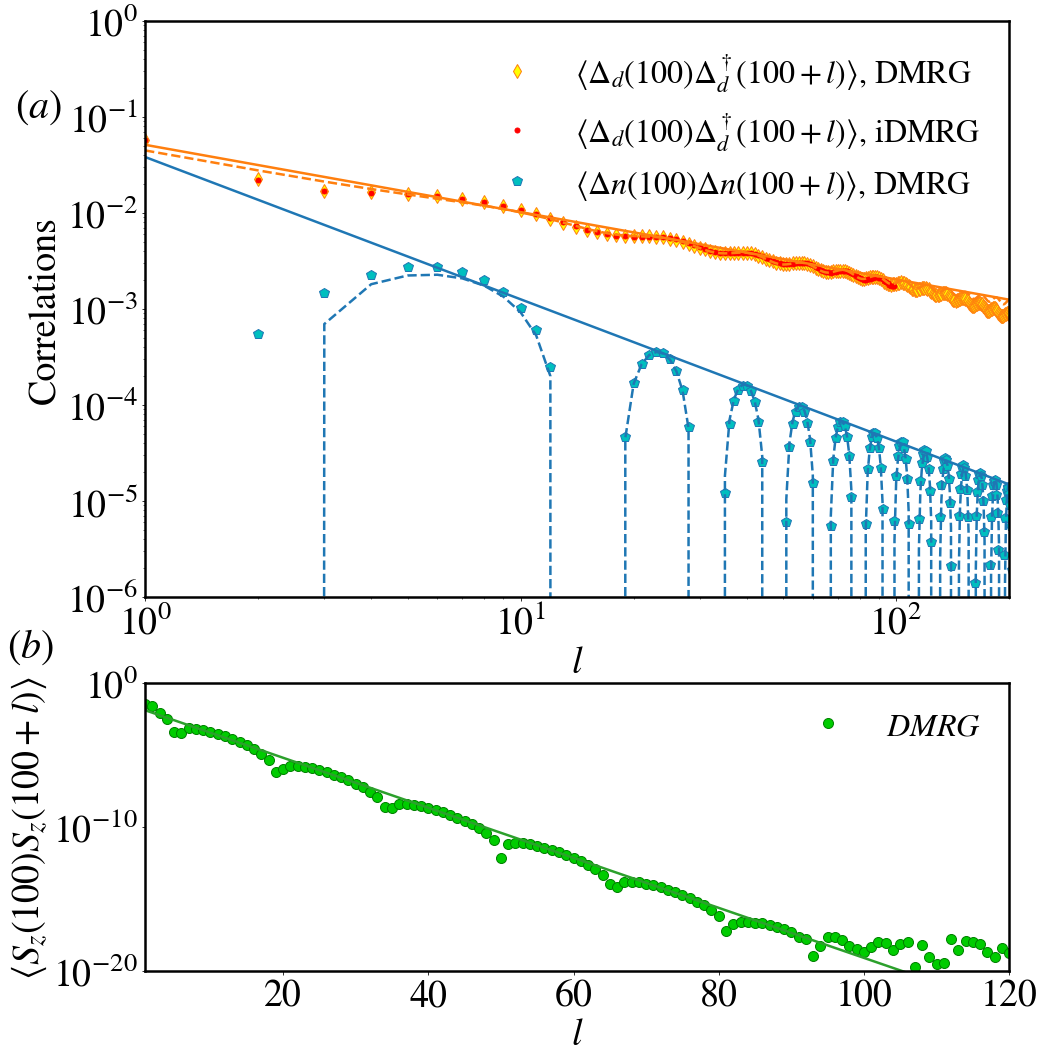}
\caption{\label{fig:corr_dmrg}(a) The unconventional pair correlations and density correlations for a repulsive Hubbard ladder (Eq.~\ref{eq:Ham_eff}) with $U=8J_\parallel, J_\perp^\text{eff}=1.4 J_\parallel$ at filling $n=0.9375$ calculated with DMRG for a ladder of size $384\times 2$  using a bond dimension of up to $M=4000$ and a truncation error $E=10^{-8}$ and with iDMRG for unit cell of size $32\times 2$  using a bond dimension of up to $M=5000$ and a truncation error $E=10^{-12}$. The correlations show an algebraic decay with additional oscillations and are fitted (dashed lines) by $l^{-\nu}\left(a+b\cos(2\pi n (l-l_0))\right)$  in order to extract the exponents where $\nu,a,b$ and $l_0$ are the fitting parameters. The extracted exponents are $K_d=1.43$ (DMRG), $K_d=1.42$ (iDMRG) and $K_n=1.48$ (DMRG) which identify the dominating d-wave superconductivity.  As a guide to the eye,  the fitted algebraic decay without oscillations is shown with solid lines. (b) The exponential decay of the spin correlations.} 
\end{figure}

Fig.~\ref{fig:Kd_idmrg} shows the ground state phase diagram of the system for repulsive interaction $U=8J_\parallel$ and filling $n=\frac{N_\uparrow+N_\downarrow}{2L}=0.9375$, zero magnetization $M_z=N_\uparrow-N_\downarrow=0$ versus the ratio of the tunnelling amplitudes $J_\perp^\text{eff}/J_\parallel$. At small ratios of the tunneling amplitudes $ J_\perp^\text{eff}/J_\parallel\lessapprox 0.9$, the charge-density-wave state is dominant. At intermediate ratios the unconventional superconducting state is stable. At large ratios $ J_\perp^\text{eff}/J_\parallel\gtrapprox 1.65$ the state reenters the charge-density-wave state. 

In order to extract the state diagram (Fig.~\ref{fig:Kd_idmrg}), the unconventional-pair exponent $K_d$ is calculated for different ratios of the effective tunneling amplitude using the infinite DMRG (iDMRG) method. For the parameters  $J_\perp^{\text {eff}}=1.4J_\parallel$, $U = 8 J_\parallel$, the corresponding correlations and fits are shown in Fig.~\ref{fig:corr_dmrg} and compared to finite system calculations obtained by DMRG in order to check their accuracy. We use a high-performance density matrix renormalization group  code for large finite systems (DMRG) and for infinite system (iDMRG) with ITensor \cite{itensor} which enables us to target correlations over longer distances than previously obtained \cite{DolfiTroyer2015}.  We checked the convergence of the obtained correlations to a sufficient accuracy.

We find that our results can be fitted in a region of distances up to $l=100$ (for density correlations the scaling behavior is preserved up to $l=200$) with the function $l^{-\nu}\left[a+b\cos(2\pi n (l-l_0))\right]$ where $\nu,a,b$ and $l_0$ are the fitting parameters. On top of the power-law decay an oscillation is observed at a wave vector $4k_F=2\pi n$. The extracted value of exponents from density and pair correlations for $L=394$ are $K_n = 1.48 $ and $K_d=1.43$ and are already close in their value. From the numerics (not shown) we observe that for smaller size of the system $L<394$ the power-law exponent extracted from density correlation lies above this value $(K_n> 1.48)$ and the exponent extracted from unconventional pair correlation lies below it $(K_d < 1.43)$. For example we obtain $K_n =1.52 $ and $k_d=1.4$ for system of size $L=192$ which is the same value as calculated in the previous work \cite{DolfiTroyer2015}.  Thus, the results for different system sizes suggest, that the values of $K_n$ and $K_d$ become equal in the thermodynamic limit of $L\to \infty$ and that the resulting Luttinger liquid parameter $K_\rho$ lies in the interval of $K_\rho \in [1.43, 1.48]$. The value of the iDMRG calculations lies already very close to that value. Since the iDMRG calculations are computationally much less costly, we use these to extract the approximate crossover between the charge density-wave and the superconducting phases (Fig.~\ref{fig:Kd_idmrg}).
 
The main aim of this work is to dynamically enhance unconventional superconducting correlations by driving effectively the system from a charge density wave to a state which resembles the superconducting state (see the arrow in Fig.~\ref{fig:Kd_idmrg}). To gain an insight into the effect of the proposed time-dependent modulation of the energy offsets (Eq.~\ref{eq:Ham_t}), we use a Floquet description \cite{Floquet1883} and assume $A(t)$ to be constant. In the high frequency limit $\hbar \omega \gg J_\parallel, J_\perp$ considering a perturbative expansion in powers of $\frac{A}{\hbar\omega}$ are given by
\begin{eqnarray}\label{eq:Ham_eff}
  &&H_{\text {eff}}^{(1)} = H_0(J_\perp^\textrm{eff}).
\end{eqnarray}
Within this approach the time-dependent driving results mainly in a modified effective rung tunneling amplitude. The rung tunneling is modified as $J_\perp^{\text {eff}}=J_\perp {\cal J}_0 (\frac{2A}{\hbar \omega})$ where ${\cal J}_0$ is the Bessel function of order zero.

This means that we can start initially with a material which has a CDW as the ground state and then switch the driving to a value such that $J_\perp^{\text {eff}}/J_\parallel$ corresponds to an unconventional superconductor (see the arrow in Fig.~\ref{fig:Kd_idmrg}). Thus, within the effective approach the formation of the unconventional superconductor can be induced dynamically.

In order to study whether this superconducting state predicted by the perturbative effective model can be reached, we determine the full time-evolution of the time-dependent Hamiltonian (Eq.~\ref{eq:Ham_t}) using the time-dependent MPS (tMPS) method for finite systems \cite{WhiteFeiguin2004,DaleyVidal2004}. Starting from a charge-density-wave dominated ground state of the system with Hamiltonian $H_0$ (see Eq.~\ref{eq:Ham_t}) and parameters $J_\perp=3.9598 J_\parallel, U=8J_\parallel$ for a ladder of size $32\times 2$ and filling $n=0.9375$, the modulation is switched on at time $t=0$. In order to model a slow switch on of the driving as it can occur via the excitation of phononic modes \cite{SubediGeorges2013}, the modulation amplitude $A$ is ramped up linearly in time with the ramp time $T_\text{ramp}$ until it reaches the maximum value, i.e.~$A(t)= A_f t/T_\text{ramp} $ for $t<T_\text{ramp}$ and $A(t)=A_f$ for $t>T_\text{ramp}$. 
We choose the driving parameters $\hbar\omega = 100 J_\parallel$ and $A_f=88.8 J_\parallel$ which give the effective tunneling $J_\perp^{\text{eff}}=1.4 J_\parallel$. The corresponding ground state is an unconventional superconductor with almost maximal superconducting correlations, i.e.~the maximum value of the Luttinger liquid parameter $K_\rho= 1.45 \pm 0.3$ (see Fig.~\ref{fig:Kd_idmrg}). 

\begin{figure}
\includegraphics[width=0.99\linewidth]{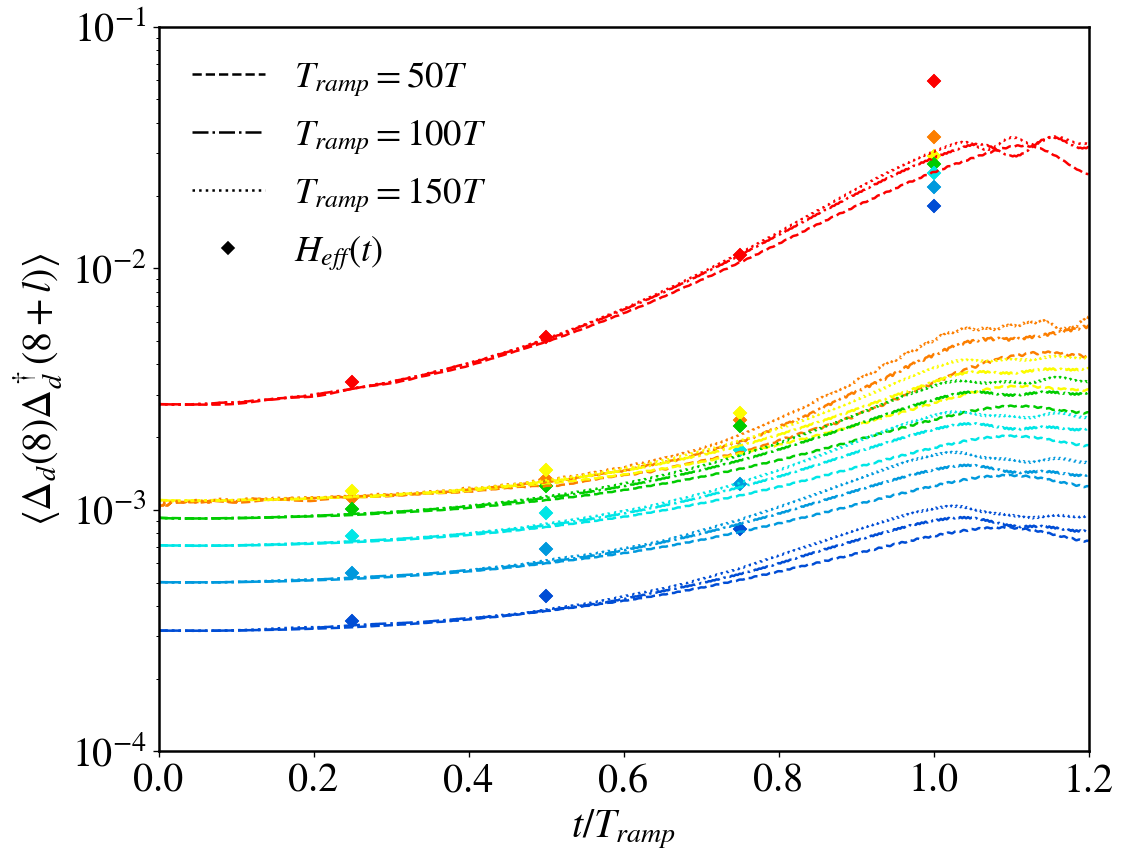}
\caption{\label{fig:DxDx_t} The dynamics of unconventional pair correlations at different distances $l=1$ to $l=7$ (from up to down) for different ramp times. The unconventional pair correlations increase during the ramp time and later oscillate around an average value. The system is $32\times 2$ ladder with filling $n=0.9375$. The parameters are $J_\perp=3.9598 J_\parallel, U=8J_\parallel, A_f=88.8 J_\parallel$. The frequency of periodic driving is $\hbar\omega= 100 J_\parallel$ with $T=\frac{2\pi}{\omega}$. Symbols mark the corresponding effective ground states.  } 
\end{figure}

 In Fig.~\ref{fig:DxDx_t} the results for the evolution of the unconventional pair correlations during the switch on of the drive is plotted for a time step $dt=0.01 T$ with $T=2\pi/\omega$ and a bond dimension $M=1000$. The convergence of these results to a sufficient accuracy is checked for different bond dimensions $M=1000,\,1500,\,2000$ and different times steps $dt= 0.01T,\,0.005 T,\,0.001T$. The unconventional pair correlations show for all considered distances and ramp times a large rise on the time scale of the ramp time. After the ramp is concluded, the pair correlations saturate and oscillate around a saturation value. The saturation value for the short range correlations lies approximately one order of magnitude above the initial value. This shows that an enormous enhancement of the unconventional pair correlations can be induced dynamically by the external drive. 

 We compare the results from the exact time-evolution with the predicted ground state results from the effective Hamiltonian Eq.~\ref{eq:Ham_eff} at different times of the ramp. Considering $J_\perp^{\text {eff}}=J_\perp {\cal J}_0 (\frac{2A(t^*)}{\hbar \omega})$ for a few chosen $t^*$ during the ramp and the final time, we calculate the unconventional pair correlation for the corresponding effective ground state shown as dots in Fig.~\ref{fig:DxDx_t}. We see that the short distance correlations $l=1$ follows approximately the expected values of the effective model at the initial times $t/T_{ramp}<0.8$. At larger ramp times $t\approx T_{ramp}$, larger deviations can be observed. The time-evolved correlation remains below the predicted value of the effective model. This effect becomes even more pronounced for larger distances ($l>1$) of the pair correlations for which the time-evolved value of the correlation remains already at earlier times below the predicted effective value.

 Comparing different ramp times, the results for the correlations with slower ramp times reach larger values and come closer to the predicted effective value. The results for different ramp times start to deviate approximately at the time when deviations of the effective model occur and, thus, we attribute part of this deviation to the non-adiabatic following during the switch on of the driving (a similar effect had been seen in \cite{PolettiKollath2011} for bosonic atoms subjected to a Floquet drive). Additionally, the first order effective Hamiltonian $H_\text{eff}^{(1)}$ might not be accurate enough for the chosen parameters and some deviations might stem from this. 

\begin{figure}
\includegraphics[width=0.99\linewidth]{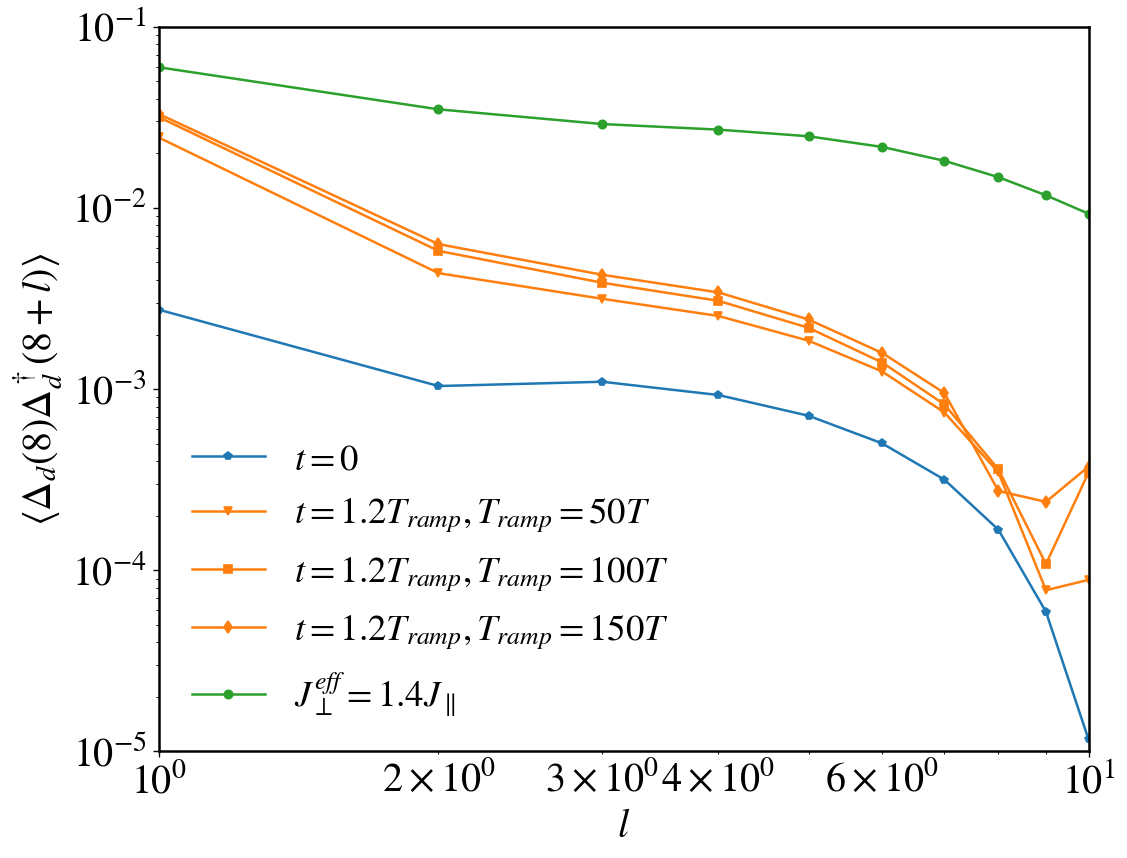}
\caption{\label{fig:DxDx_x} The unconventional pair correlations at initial time $t=0$ and at the time $t=1.2 T_{\text{ramp}}$ for three different ramp times $T_\text{ramp}= 50T, 100T, 150T$. The correlations in the ground state of the effective Hamiltonian with $J_\text{eff}=1.4 J_\parallel$ are plotted for the comparison. Parameters are the same as in Fig.\ref{fig:DxDx_t}.} 
\end{figure}

In order to investigate in more detail the final value reached after the switch on of the driving, in Fig.~\ref{fig:DxDx_x} the pair correlation after the ramp time ($t=1.2 T_\text{ramp}$) is plotted versus distance $l$ for the different ramp times $T_\text{ramp}=50T, 100T, 150T$ and compared to the initial correlations and the predicted final value of the effective model \footnote{Let us note that the algebraic decay predicted by bosonization for the ground states in only evident in larger systems.}. A significant dynamic enhancement of the unconventional pair correlations by approximately one order of magnitude for $l=1,2$ is evident. Even though the high value of the final effective correlations is not fully reached, the enhancement increases with slower ramp times. We expect from this finding, that slower ramp times could be utilized in order to enhance the values of the unconventional pair correlations even more. 


In this work we investigated the dynamic enhancement of unconventional-pair correlations in a periodically modulated Hubbard ladder. Starting with a state which is initially in a charge density wave state, the drive induces unconventional-pair correlations and drives the system towards an unconventional superconducting phase. We used the exact t-DMRG methods in order to determine the dynamics. Deviations from the predicted ground state of an effective Floquet-picture are found which are mainly attributed to the non-adiabatic switch on of the perturbation. Our results give a proof of principle that unconventional superconducting correlations can be induced via a dynamic control. Our results could be probed in quantum gas experiments in optical superlattices, where a dimerized modulation of an energy offset has already been realized \cite{EckardtHolthaus2005, LignierArimondo2007, SiasArimondo2008, AidelsburgerBloch2011,  AidelsburgerGoldman2015}. A careful analysis of different phononic modes for example organic structure which realize Hubbard ladders in order to identify a mechanism for the realization in such systems and following simulations which take realistic parameters into account would be very fruitful. 

We thank J.S.~Bernier, T.~Giamarchi, M.~K\"ohl and E.~Orignac for enlighting discussions.
We acknowledge support from DFG (Project number 277625399- TRR 185, project B4, and CRC 1238 project number 277146847 - projects C05, and Einzelantrag) and the ERC Phonton (Grant Number 648166). A.~S.~thanks the research council of Shahid Beheshti University, G.C.~(project number SAD/600/655)

\end{document}